# Sparsity for Ultrafast Material Identification

Yurui Qu, Qingyi Zhou, Jin Xiang, Zongfu Yu*

*Department of Electrical and Computer Engineering, University of Wisconsin–Madison, Madison, WI 53705, USA*

*Corresponding Author: zyu54@wisc.edu

Abstract

Mid-infrared spectroscopy is often used to identify material. Thousands of spectral points are measured in a time-consuming process using expensive table-top instrument. However, material identification is a sparse problem, which in theory could be solved with just a few measurements. Here we exploit the sparsity of the problem and develop an ultra-fast, portable, and inexpensive method to identify materials. In a single-shot, a mid-infrared camera can identify materials based on their spectroscopic signatures. This method does not require prior calibration, making it robust and versatile in handling a broad range of materials.

Material identification is important to a broad range of applications such as chemical synthesis [1, 2], food analysis [3-5], art evaluation [6, 7], forensics [8] and biosensing [9-11]. A typical method for material identification is mid-infrared spectroscopy [12-16]. Mid-infrared spectra contain absorption peaks that provide a wealth of information about specific functional groups. The region from 500 to 1500 cm$^{-1}$ is known as fingerprint region because the spectrum in this region is often unique for different compounds. In contrast, many compounds with distinct mid-infrared features often lack visible spectral features. For example, the compounds PAN (Polyacrylonitrile), PMMA (Polymethyl Methacrylate), PVP (Polyvinylpyrrolidone) are indistinguishable in visible light while their mid-infrared spectra are very different (Fig. 1a). The most widely used method of measuring the mid-infrared spectra is to use Fourier transform infrared spectroscopy (FTIR). Such measurement consists of thousands of measurement points in a scanning setup. Acquiring these spectra is time-consuming and requires expensive and bulky instrument. In many scenarios, fast measurement speed are needed, such as explosives or hazardous chemical detection [17, 18] and field environmental monitoring [19, 20]. Ultrafast material identification has been challenging using traditional mid-infrared spectroscopy.

In practical applications, we often need to identify one material out of a limited number of potential candidates. Therefore, this identification problem is fundamentally sparse [21]. It was shown that compressive sensing can use an extremely small number of measurements to solve sparse problems [22, 23]. For example, to identify one out of 1000 options, on average we only need to perform about 10 measurements [24]. By taking advantage of the sparsity of the problem, we could substantially reduce the number of measurements and thus make material identification fast, inexpensive, and portable [25-47].

We now illustrate the idea using a specific example. Suppose we need to identify one out of 1000 materials. We can represent the signal using a material identification (ID) vector $x$ with 1000 entries ($x \in \mathbb{R}^m, m = 1000$). One of the entries in the vector $x$ is non-zero, signaling the presence of the corresponding material while other zero elements represent the absence of those materials as shown in Fig. 1b. This signal is very sparse in the material space.

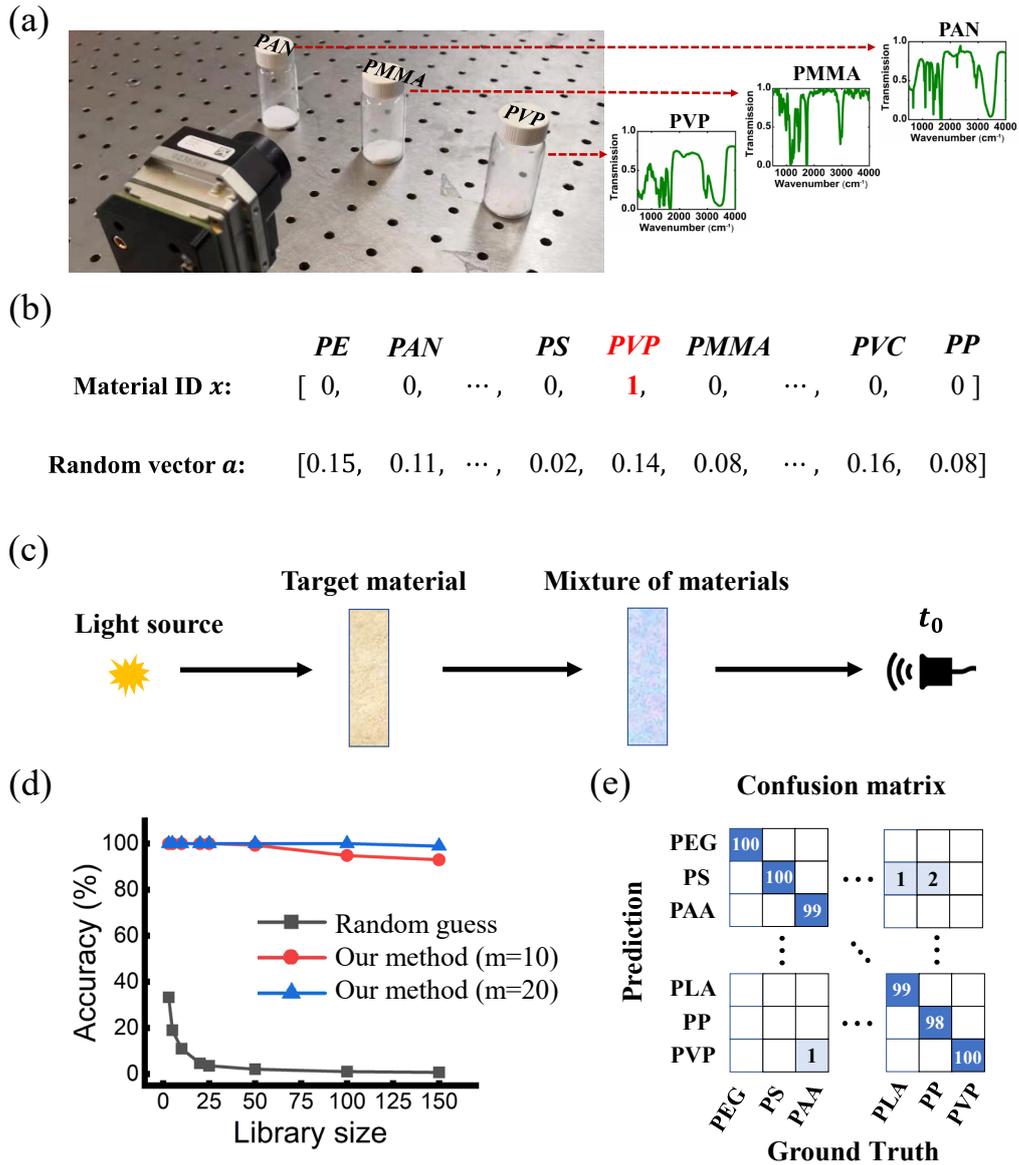

Figure 1. (a) Three materials (PAN, PMMA, and PVP) were identified using a mid-infrared sensor. Three compounds are indistinguishable in visible light, but can be identified in the mid-infrared wavelength range. (b) A specific example of sparse target signal $x$ and random vector $a$. (c) The incident light is transmitted through the target material and a mixture of materials. The transmitted light detected by a detector is $t_0$. (d) Identification accuracy at different library sizes. The task is performed by random guess and by our method (number of measurements $m=10$ and $m=20$), respectively. (e) The partial confusion matrix of the identification results. The material library size is 150 and the number of measurements is 20.

The sparse nature of the problem makes it possible to achieve fast material identification using compressive sensing. Compressive sensing starts by projecting the unknown signal $x$ to a random vector $a$, which is also defined in the same material space. Figure 1b shows an example vector $a = \{0.15, 0.11, \cdots, 0.02, 0.14, 0.08, \cdots, 0.16, 0.08\}$, in which each element can represent the volume percentage of each material. The summation of all elements in vector $a$ is one. The projection yields one measurement: $a \cdot x$. Such projection is abstract, and we will discuss its physical implementation later. We can perform this projection for $n$ times with different random vectors. We can put all $n$ measurements together as

$$b = Ax, \qquad (1)$$

where $A = [a_1, a_2, \cdots a_n]$ and $b = [b_1, b_2, \cdots b_n]$. If the unknown signal $x$ is sparse, the above equation can be solved by minimizing $\ell_1$ norm [22]:

$$min\|x\|_1 \; subject \; to \; b = Ax. \qquad (2)$$

The number of measurements needed depends on its sparsity and dimensionality. For one out of 1000, it usually requires about 10 measurements. It is well documented that such problem can be solved with accurate identification [21] (see example in Supplementary Fig. S1).

The analysis above shows that it is possible to rapidly identify material using just a few projections, far less than today's prevailing methods such as spectroscopy, which often involves thousands of measurement points. However, we still need to map the abstract operation $a \cdot x$ in Eq. (1) to a physical measurement. We need to transform Eq. (1) into similar projections realizable in certain physical domain $a^p \cdot x^p$, where we use superscript $p$ to denote the vectors in certain physical domain.

There are many ways to perform the transformation, which would correspond to different physical measurements. Next, we show one based on a setup shown in Fig. 1c. The target material to be identified is shown by light yellow color. For the projection with a random vector $a$, we use a thin film of a mixture of materials shown by blue color. It is mixed according to the material ratio defined by the elements in the random vector $a$. A broadband light source illuminates the target material and then the mixture. The transmission $t_0$ is recorded by a photodetector (Fig. 1c). Separately, we also perform similar transmission measurement for the target material and the mixture of materials alone as $t_1$ and $t_2$, respectively. We show in the Supplementary Materials section 2 that these measurements are equivalent to performing a projection as follows

$$a^p \cdot x^p = t_0 - t_1 - t_2 + 1, \quad (3)$$

where $a^p$ and $x^p$ represent the absorbance spectra of the mixture ($Fa$) and the target material ($Fx$), respectively. This equation is related to the material ID vector by $t_0 - t_1 - t_2 + 1 = \langle Fa, Fx \rangle$, where $\langle \cdot, \cdot \rangle$ represents inner product. The transformation matrix $F = [F_1 \ F_2 \ \cdots \ F_m]$ is the matrix of the absorbance of materials in the library. Here $F_1 \cdots F_k$ are column vectors representing the absorbance of each materials. $a$ is the volume percentage of materials in the random mixture. The matrix $F$ can be taken directly from the libraries of material properties, such as the open-source dataset from National Institute of Standards and Technology (NIST) [48].

Performing such measurement a few times using different random mixing materials, we can have a set of projections that are related to Eq. (1) by $A^p \cdot x^p = A F^T F x$, where $F^T F$ is a near-identity matrix (see details in the Supplementary Materials section 2). As an example, to demonstrate the feasibility, we randomly identify one target material from a library of material candidates. Here we use $m$ =10 projections for each identification task and perform the task 100 times to obtain statistical significance. Figure 1d shows the performance for different sizes of the library. The performance remains high even when there are 150 material candidates. On the other hand, the success rate of random guess (grey curve in Fig. 1d) quickly decreases as the library size goes up. If we have more projections, e.g. m = 20 (blue line in Fig. 1d), the performance can be further improved. Figure 1e shows the confusion matrix, indicating that materials are correctly identified most of the time.

Next, we discuss how to implement ultrafast material identification in a practical system. It is very helpful to use material mixtures to conceptually understand the projection in Eq. (1). However, it is not a scalable approach for practical applications. First, mixing many different materials is not readily compatible with scalable semiconductor fabrication. Secondly, the material to be identified and its competing candidate materials change from one application to another. We would like to have one device that can be used for many applications. To address this issue, the random spectral projection, which is the key to compressive sensing, is offered by the random combination of materials. Similar randomness can also be achieved by using structure-enabled spectral encoders (see the Supplementary Materials section 3 for detailed discussion). Figure 2c shows representative spectra from random mixture of materials, and structure-enabled encoders. These structures can be defined easily in semiconductor fabrication process. More importantly, the same device can be used to identify arbitrary materials without any prior calibration for the materials or competing material options.

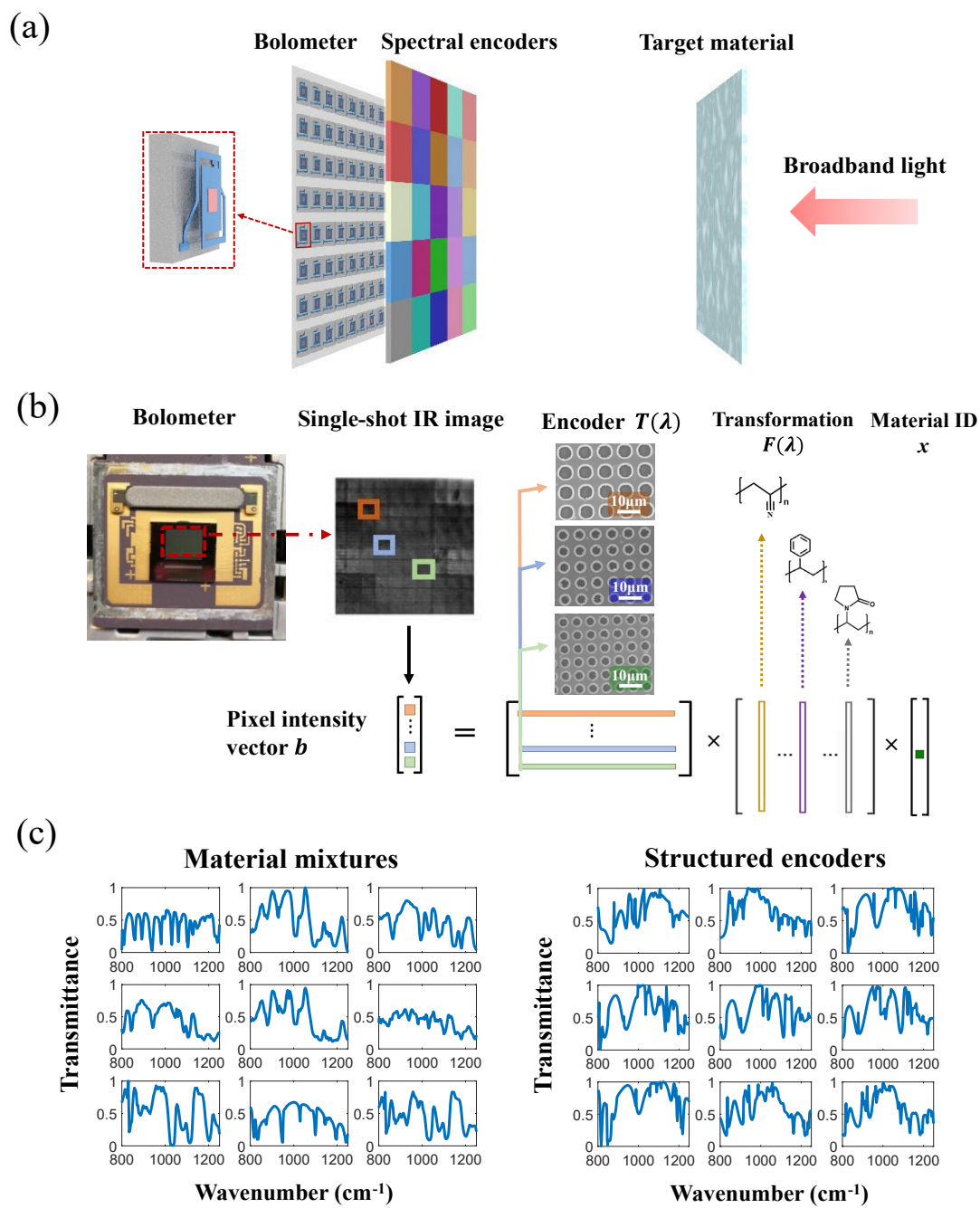

Figure 2. (a) Schematic of the system for rapid material identification. Arrays of spectral encoders with diverse mid-infrared spectral features are integrated on top of arrays of bolometers. Light passes through the target materials coming from the right. (b) Photo of the bolometer and SEM images of microstructure spectral encoders. The algorithm

of material identification is also shown the diagram. The pixel intensity $b$ is extracted from a single-shot mid-infrared image captured by the bolometers. The SEM images of three encoders are shown. (c) The transmission spectra of material mixtures (left) and the structured encoders (right) both show the randomness needed to achieve compressive sensing.

Figure 2a shows an implementation where we integrated a set of spectral encoders on top of bolometer arrays (FLIR Tau 2). Each encoder together with the bolometer behind it perform one projection. All encoders are different. Having arrays of encoders and bolometers allows us to measure multiple projection in a single shot (Fig. 2b). The spectral encoders replace the material mixture used in Fig. 1c. The spectral encoders are thin films made from germanium (Ge) patterned with a unique microscopic structure to produce a unique spectral signature (see details in the Supplementary Materials section 4 and 5). The thickness of the Ge film is 3 μm. The microscopic structures were fabricated by photolithography and inductively coupled plasma etching. Figure 2c shows their transmission spectra in comparison to the spectra of material mixtures. They both provide rich and random spectral features in the mid-infrared wavelength.

Next, we experimentally demonstrate rapid material identification of 150 materials picked from a public dataset provided by NIST [48]. It is important to note that the material library $F(\lambda)$ is directly constructed using the spectra reported in the database (see examples in the Supplementary Materials section 6). We do not need to measure or calibrate their spectra. Any new unknown material can be identified without prior measurement and calibration, as long as the spectra of the suspect material and competing candidates are available in the literature. All measurements are obtained within 20 ms.

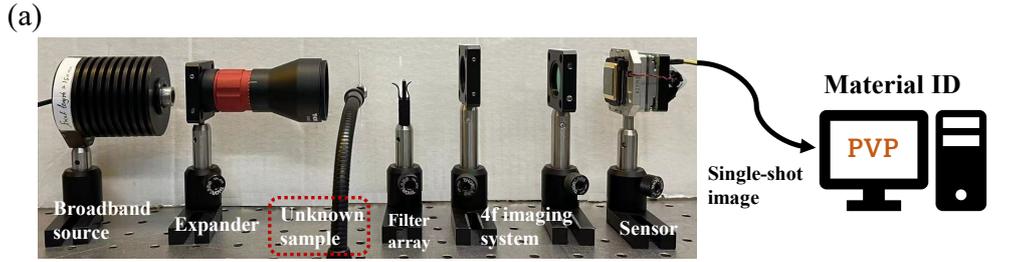

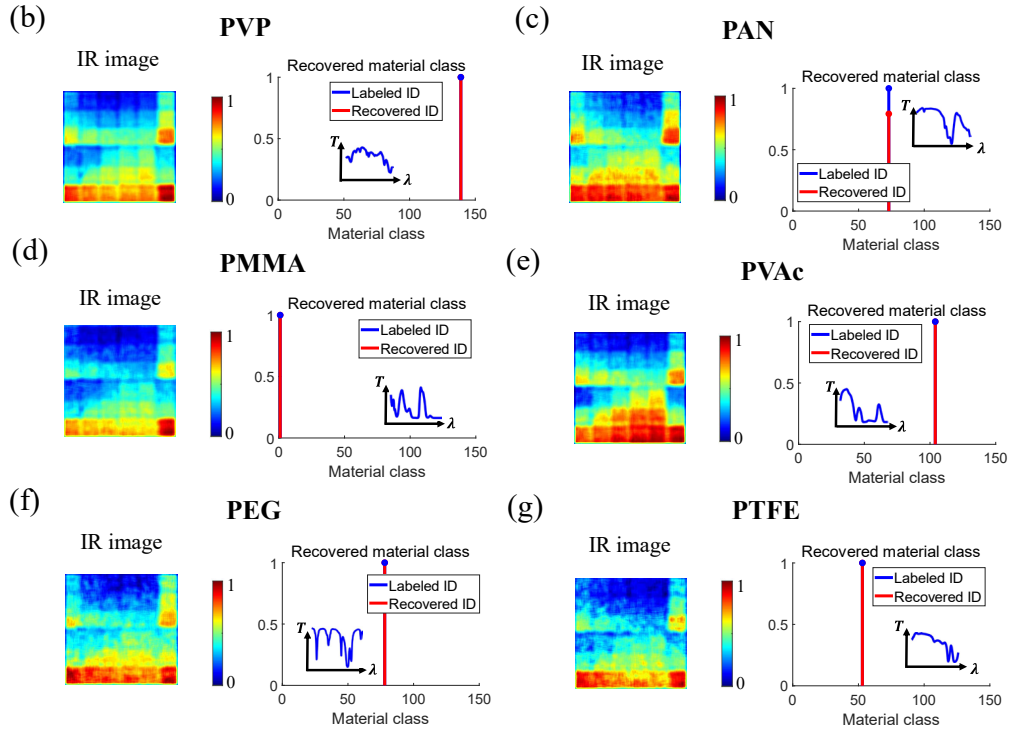

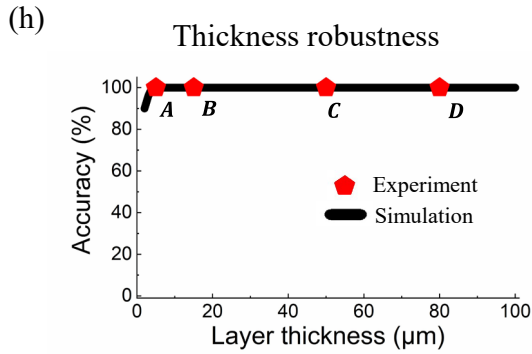

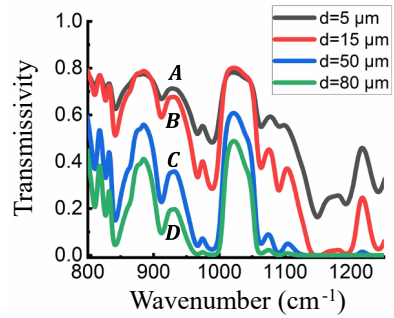

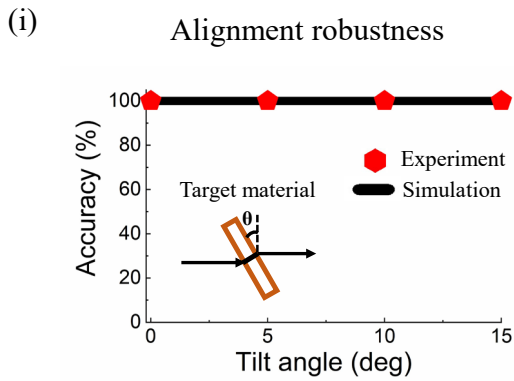

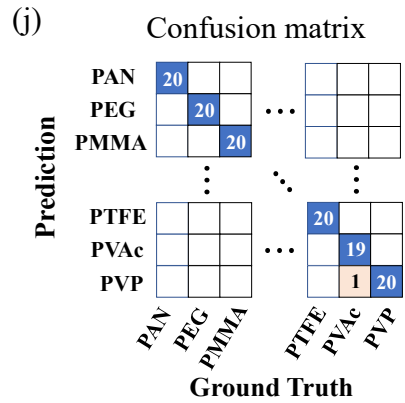

Figure 3. Experimental setup and results. (a) The experimental setup for material identification. (b-g) The results of experimental identification of different target materials. Left: snapshot mid-infrared image of each material. The pixel values are normalized to a range between 0 and 1. Right: recovered material ID and labeled ID (ground truth). The insets in the right panel are the spectra of the corresponding materials. (h) The robustness analysis of layer thickness variations. Identification accuracies were measured and calculated for different layer thicknesses. The measured spectra of four thicknesses are plotted in the right panel. (i) The robustness analysis of alignment. Identification accuracies were measured and calculated from 0° to 15° tilt angle. (j) The partial confusion matrix of experimental results.

Figure. 3a shows the experimental setup, which consists of a blackbody as the light source, a 10-fold beam expander, spectral encoders, a 4f imaging system and a mid-infrared bolometer array. Our device contains 36 spectral encoders, arranged into a 6 × 6 array, each with a size of 300 × 300 μm$^2$. The spectral response range of the bolometer array (FLIR Tau 2) is from 7.5 to 13.5 μm (1333 to 740 cm$^{-1}$). Details of the experimental setup are included in the Supplementary Materials section 7.

To test the instrument, some common materials are selected for the identification task such as polyvinylpyrrolidone (PVP), polyacrylonitrile (PAN), poly(methyl methacrylate) (PMMA), polyvinyl acetate (PVAc), polyethylene glycol (PEG) and polytetrafluoroethylene (PTFE). These materials are selected for no particular purpose, but are simply common and easily accessible in the laboratory. It should be clarified that our instrument can identify a wider range of materials, not limited to those for the purpose of demonstration. These target samples are prepared by spin-coating on a potassium chloride (KCl) substrate.

The identification results are plotted in Figs. 3b-g. The left panels of Figure 3b-g show the snapshot infrared images of the target materials captured by the bolometer array. Each captured image displays a unique pattern. The corresponding material ID vector is reconstructed use $\ell_1$-minimization to yield the sparsest solution $x$ (see details of the algorithm in the Supplementary Materials section 8). The reconstructed material IDs are consistent with the labeled IDs, indicating that each target material is correctly identified (right panels of Figs. 3b-g). The insets show the corresponding spectra.

The material identification method described above has excellent tolerance over the variation of sample thickness and alignment. We first show why our method is highly tolerant to thickness variation. The signal for a light source illuminating the encoder alone without the target material is recorded as $b_0$. The signal for the light source illuminating both the target material and the encoder is recorded as $b_1$. By measuring $b_0 - b_1$, a system of linear equations for the measurement can be obtained, where the thickness of the target material $d$ is only a scale factor. The measurement can be written as

$$b_0 - b_1 = d \cdot A\alpha, \tag{4}$$

where $A$ is the encoder matrix and $\alpha$ is the absorption coefficient of the target material. The index of the non-zero entries in the sparsest solution $x^*$ indicates the reconstructed material ID. Using equation (4), we can prove that, when the thickness $d$ changes, the indexes of the non-zero entries in $x^*$ do not change (see detailed proof in the Supplementary Materials section 9). Therefore, the recognition accuracy of our method can remain high over a wide layer thickness range.

The performance is evaluated in both simulation and experiment for different layer thicknesses. Without loss of generality, the commonly used polymer PMMA was chosen. In experiment, each thickness was measured 50 times independently and then averaged. Figure 3h shows the recognition accuracies for different layer thicknesses. The recognition accuracy remains high when the layer thicknesses are in a very wide range between 2 μm and 100 μm. This shows that our method is robust against sample thickness variations. The transmission spectra of the target material with different layer thicknesses (5 μm, 15 μm, 50 μm, 80 μm) are plotted in the right panels of Fig. 3h. Their spectra are different, but retain unique spectral features.

Next, we discuss alignment tolerance. A typical misalignment is the tilt of the sample. At different tilt angles, optical path length changes. Since our algorithm is robust to the thickness variation, it is easy to prove that the algorithm is also robust to the optical path length. Both experimental and theoretical results show that our method is robust to the tilt angle from 0 to 15° (Fig. 3i). Each angle is also measured 50 times and then averaged. To obtain the statistical significance of the experimental results, we performed each identification task multiple times. The results are plotted in the confusion matrix (Fig. 3j). The results show that our method is very accurate and stable.

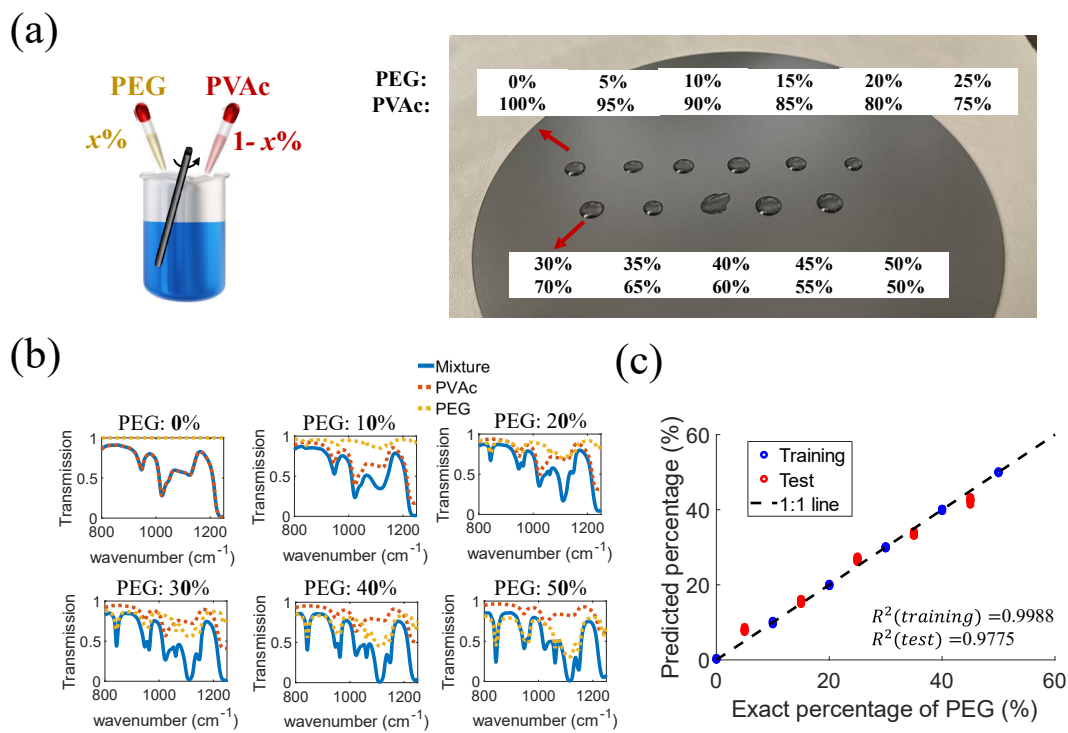

Figure 4. The measurement of the material ratio in a mixture. (a) Left: two materials, PEG and PVAc, were mixed in certain volume ratio. Right: the optical images of the mixture. Two materials were mixed in 11 different volume ratios. (b) The spectra of the mixtures (blue line) with different volume ratios were measured using FTIR. To clearly show that the spectral variation of the mixture is cause by the different ratios of the two materials, we also plotted the spectra of pure PEG (yellow dashed line) and PVAc (red dashed line) multiplied by different scale factors. (c) Experimental results: predicted vs. exact volume ratio of the two materials.

Sparsity can be exploited for other applications related to material identification. We describe an approach to identify the presence of chemical bonds in the Supplementary Materials section 10 and 11.

In addition to material identification, in some cases it is necessary to know the ratios of various materials in a mixture, such as chemical production, food and environmental monitoring. In the previous section of material identification, we do not need to calibrate or train multiple times before measurement. Here we show that with a bit of training, we can expand the utility of this instrument to measure the material ratios in a mixture. As a proof of concept, we show an example of measuring the volume ratio of two materials (PEG and PVAc) in a mixture. The experimental setup used for the ratio measurement is the same as the one in Figure 3a.

We prepared 11 mixed solutions with different volume ratios of PEG and PVAc (Fig. 4a). The transmission spectra (blue lines) of the mixed materials with different volume ratios were measured (Fig. 4b). Obviously, different mixing ratios can lead to the spectral variation of the mixture. Six of the 11 material ratios were used as training data, and the other five were used as test data. The prediction performance was evaluated using the coefficient of determination ($R^2$). The prediction results $R^2$ for the training and test data are 0.9988 and 0.9775, respectively, indicating that our method can predict the material ratio very accurately (Fig. 4c). The prediction model is built using partial least squares method. Details can be found in the section 12 of the Supplementary Materials.

In conclusion, we proposed an ultrafast method for material identification. To significantly improve the speed, we exploit the sparsity of the problem to reduce the number of measurements. The same device can also be used to measure the ratio of materials in a mixture. In the future, it is possible to fully integrate the photonic encoders on bolometer arrays to enable handheld material identification devices. In addition, certain applications can be combined with deep learning algorithms [46, 47]. Such instruments could be inexpensively deployed in large quantities to accelerate the automation in material and industrial applications.